# Class Association Rules Mining based Rough Set Method

Thabet Slimani [#1]

[#] Computer Science Department, Taif University
College of Computer Science and Information Technology
[1] thabet.slimani@gmail.com
Taif, Saudia Arabia

*Abstract*— This paper investigates the mining of class association rules with rough set approach. In data mining, an association occurs between two set of elements when one element set happen together with another. A class association rule set (CARs) is a subset of association rules with classes specified as their consequences. We present an efficient algorithm for mining the finest class rule set inspired form Apriori algorithm, where the support and confidence are computed based on the elementary set of lower approximation included in the property of rough set theory. Our proposed approach has been shown very effective, where the rough set approach for class association discovery is much simpler than the classic association method.

Data Mining, RST, CAR, ARM, NAR, Bitmap, class association rules, Rough Set Theory

## I.  Introduction

Data Mining (DM) is a modern area of research very useful in computer science. The objective of DM is to extract various models of interesting, hidden, and potentially useful knowledge from databases, where the volume of collected data is huge. Knowledge exploited by data mining can be represented as rules, customs, patterns, trends, etc. DM [1] is a prominent tool which encloses several techniques: Association, Clustering, Classification and Deviation. Association rule mining (ARM) [2] is defined to extract the important correlation and relation included in large amount of data. Association rule mining aims to find interesting relationships from the data in the form of rules. ARM, are originally applied on market basket analysis seeking to study the buying habits of customers [3]. Interesting association rules discovery can be used to help the decision making process.

As a formal definition, an association rule is a relation in the form of implication A➔B between two disjunctive sets of items A and B. A typical example of an association rule on "market basket data" is that "80% of customers who purchase spaghetti also purchase sauces ".

Two quality measurements characterize each rule, support and confidence.

The expression if *A* then *B (A➔B)* is a regular association rule *for* attribute sets A and *B (*with some *confidence)*. Consequently, an association rule A➔B is regular means that if *A* maximally then *B* maximally [4]. More deeply, the rule A➔B has confidence CF if CF% of transactions in the set of transactions D that contains A also contains B. The rule A➔B has support SP if SP% of transactions in D contains A∪B. To find regular association rules is a problem to find all association rules having a support and a confidence greater than the threshold of minimum support specified by an expert (called MinS) and threshold of minimum confidence (called MinC ) respectively.

Additionally, ARM can be exploited in information retrieval where there exist a need to identify association between keywords.

Different types of association rules can be enumerated: rules-based types of values handled, rules-based levels of abstraction handled and rules-based dimensions of data involved. The first type can be classified into Boolean or quantitative association rules and the second type can be classified into single-level and multi-level association rules. In multidimensional database, ARM can be classified into single dimensional association rules (SDAR) and multidimensional association rules (MDAR).

A single distinct predicate with multiple occurrences is referred to us as SDAR where transactional data is used. The terminology of *single dimensional* is used to consider each distinct predicate in the rule as a dimension. More specifically, items in a rule are assumed to belong to the same transaction. For instance, in market *basket analysis*, the SDAR representation of the Boolean association rule "diapers ⇒ beer" can be written as follows [5]:

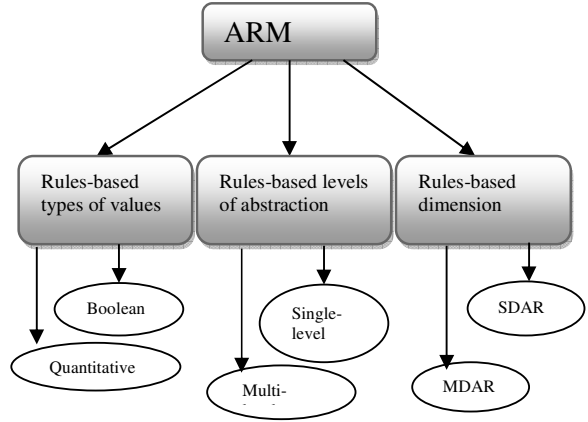

Figure1: Association Rule Mining Types Tree

**R1: buys(x, "diapers") ⇒ buys(x, "beer") [10% (supp), 70% (conf)].**
The MDAR representation uses relational data where an Attribute X in a rule is assumed to have value x, attribute Y has the value y and attribute Z has the value z in the same tuple. For instance, in market *basket analysis*, with the same example of the SDAR representation, it considers items in the rule varies from two to more dimensions or predicates, e.g. *"buys", " transaction_time", "customer_category"*. For instance, R2 is an example of MDAR:

**R2: Age(A,"20..29") ∧ income(A,"60K..80K") ⇒ buys(A, High Resolution TV)**

The Rules that concern associations between the presence or absence of items are Boolean rules: For e.g. *"buys an item A"* or *"does not buy an item A" (e.g. R3)*

**R3: buys(x, "A") ^ buys(x, "B") ⇒ buys(x, "C") [0.2%, 60%]**

The rules that concern associations between quantitative items or attributes are quantitative rules. For instance, R4 is an example of quantitative association rules:

**R4: age(x, "20..29") ^ income(x, "18..38K) ⇒ buys(x, "PC") [1%, 80%]**

Rough set theory can be used for data mining when the available information is insufficient to determine the exact value of a given set, based on lower and upper approximations for the representation of a concerned set [6]. By using this theory, it is possible to extract rules that are similar to normal associations. However, we investigate the rough set approach to discover class association rules and we show that this approach is simpler than the classic association method.

The paper is structured as follows: Section 2 describes ARM background which explains data preparation for further process with rough set approach. Additionally, it discusses the meaning of itemset, support and confidence of rule, how to transform relational schema into bitmap table and the meaning of class association rules. Section 3, presents the rough set model and its applications. Section 4 discusses how to apply RST to class association rules, how to represent data with RST and the algorithm C_Apriori adopted for CAR mining. Finally, section 5 concludes the paper.

## II. BACKGROUND

Agrawal et al. [3] is the first author that introduces Association Rule Mining that begins a well-known data mining research field. The main idea is to extract common model of mined knowledge under format of Association Rules set (ARs) based on data stored in transactional database D. Let I = {$i_1, i_2, ..., i_{n-1}, i_n$} be a set of items or database attributes, and T = {$t_1, t_2, ..., t_{m-1}, t_m$} be a set of transactions or database records, T describe D, where each $t_j \in T$ includes the items in the set I′ ⊆ I.

The implication of co-occurring relationship between two sets of items in D is what it defines an association rule. However, an association rule is expressed in the form of the implication: "antecedent (A) ⇒ consequent (B)", where A, B ⊆ I and A ∩ B=Ø. There are two ways to measure the usefulness of an association rule: objective and subjective measures. Objective measures involve two threshold values that are commonly used in ARM to measure the significance of an association rule:

- ☒ **Support:** An itemset is formed by a set of items S. The proportion of transactions T' in T for which S ⊆ T is the support of S. The rule R (A➔B) occurs with support s if s% of transaction in D contains A∪B. The rule that have a support s greater than a user-supplied support threshold (σ) is defined to be significant (have minimum support).

- ☒ **Confidence**: It is based on a user-supplied confidence threshold α, and aims to discover how "strongly" a rule antecedent A implies another rule consequent B. The association rule A➔ B occurs with confidence c if c% of the transactions in D containing A also contains B. The association rule A➔B is said to be valid if the support for the A and B co-occurrence exceeds σ, and the confidence of this association rule exceeds α.

The support is computed as follows:

$$S(A \cup B) = |A \cup B| / |T| \quad (1)$$

TABLE I
Example of Support Measure

| TID | Items | Support=Occurrence/Total Trans |
|-----|-------|-------------------------------|
| 1 | ABD | |
| 2 | AB | Total Trans=4 |
| 3 | ABC | Support({AB})=3/4=75% Support({BC})=2/4=50% |
| 4 | BCD | |

Where |A ∪ B| is the transactions number containing the set A ∪ B in T, and |T | is the cardinality of the set T.
The confidence is computed as follows:

$$C(A \Rightarrow B) = S(A \cup B) / S(A) \quad (2)$$

TABLE II
Example of Confidence Measure

| TID | Items | Given an implication X➔Y; Conf(X➔Y)=Supp(YUX)/Supp(X) |
|-----|-------|----------------------------------------------------------|
| 1 | ABD | |
| 2 | AB | Conf(A➔B)=3/3=100% |
| 3 | ABC | Conf(B➔D})=2/4=50% |
| 4 | BCD | |

Apriori algorithm is mainly the well-known ARM algorithm, developed by Agrawal and Srikant [3], which represents the basis of various subsequent ARM algorithms.

*A. Relation Table Types*

Generally, the process of association rules discovery uses a single table (relation) as a source of data that represents relations between items. Formally, a relation is a relational table R that *i*ncludes a set of tuples $(t_1, t_2, \ldots t_i, \ldots t_n)$, where $t_i$ represents the *i*-th tuple. A relation R can be either accompanied with binary domain or no-binary attributes. As an example of a relation RL1 with binary attributes: the presence of a computer item in a transaction or its absence represents its domain {sold, not sold}. An attribute $A^j$ is non-binary domain is represented by j items and $\sum_{i=1}^{n} j * i$ binary vectors such that n is the number of attributes of the non-binary domain. For example, for the better representation of a customer wealth level, we associate to the attribute

"income" the domain constituted by 3 (j=3) items {high, medium, low} defined as follows: a1 = {"high income"}, a2 = {"middle income"} and a3 = {"low income"}.

### B. Bitmap Representation

A relation or table uses as data source for ARM approach, some attributes are measurable with discrete variable as some numerical or textual values on behalf of some range. However, the form of original data representation could be changed exactly so that, each attribute in the new Bitmap table is an exact value of one item in original table, and each attribute value should be 1 or 0, expressing if it exist there is a '1', otherwise a '0' in the bitmap table[7].

Let be the example of table 3 where attributes representing data are {X}, {Y} and {Z}. The attribute X has two values {A and B} = {Account debited, Account credited}, the attribute Y has three values {C, D and E} = {low income, high income, middle income} and the attribute Z has two values {F, G} = {according loan, not according loan}. There are 7 items for the resultant Bitmap table {A, B, C, D, E, F and G}.

TABLE III
Table 3. Original Relation Data

| Tid | Account | income | According Loan |
|---|---|---|---|
| 1 | Debited | middle | yes |
| 2 | Debited | low | no |
| 3 | Debited | middle | yes |
| 4 | Debited | high | yes |
| 5 | Credited | high | no |

The conversion of original relation data as Bitmap table is figured in table 4 as follows:

TABLE IV
Table 4. Bitmap table after original data conversion.

| Tid | A | B | C | D | E | F | G |
|---|---|---|---|---|---|---|---|
| 1 | 1 | 0 | 0 | 0 | 1 | 1 | 0 |
| 2 | 1 | 0 | 1 | 0 | 0 | 0 | 1 |
| 3 | 1 | 0 | 0 | 0 | 1 | 1 | 0 |
| 4 | 1 | 0 | 0 | 1 | 0 | 1 | 0 |
| 5 | 0 | 1 | 0 | 1 | 0 | 0 | 1 |

### C. Class Association Rules (CARs)

Let be T a set of n transactions. Each transaction us labelled by a class y. The set of all items in $T$ is labelled by $I$ and the set of class labels is labelled by $Y$ where $I \cap Y = \emptyset$. A class association rule (CAR) is an implication of the form: A➔B where $A \subseteq I, and\ B \subseteq Y$. The following table give a comparison between normal association rules (NAR), denoted above by ARM, and class association rules (CAR):

TABLE V
Table 5. Comparison between NAR and CAR

|  | NAR | CAR |
|---|---|---|
| **Support** | Same support | |
| **confidence** | Same confidence | |
| **Consequent** | any item(s) | Has only single item. No item from $I$ appear as consequent |
| **condition** | any item(s) | No class label from $Y$ can appear as a rule condition |

Mining CARs is an objective to generate the complete set of CARs satisfying a user-specified minimum support and minimum confidence constraint.

## III. ROUGH SET

Rough set theory (RST) is a useful mathematical method that deals with inconsistency problems developed by Pawlak in 1982 [8].

RST is defined as an extension of the conventional set theory that supports approximations in decision Making [8]. The rough set is the approximation of a vague concept (set) by a pair of fixed concepts classifying a specified domain into disjoint categories named lower and upper approximations. The lower approximation describes the domain objects which are known with certainty to belong to the subset of interest, whereas the upper approximation describes the objects which possibly belong to the subset.

The theory of rough sets is described formally in the work of [8][9]. The concept of RST is described as follows: Let be the universe $\Omega \neq \emptyset$ a finite set of objects for that any subset $A \subseteq \Omega$ of the universe is called a concept in $\Omega$, and representing each knowledge by any family of concepts contained in $\Omega$. The family of classifications over the universe $\Omega$ refers the knowledge base over $\Omega$. The formal foundation of RST is based on the fact to consider the "universe" as a finite set. In database systems, the meaningfulness of updating sets (insert, delete and join) is interesting in several database applications.

More formally, let be R an equivalence relation over $\Omega$ such that $R \subseteq A \times A$, then the following properties should be considered:

- R is reflexive: aRa,
- R is symmetric: if aRb then bRa
- R is transitive (if aRb and bRc then aRc)

$\Omega/R$ denotes the family of equivalence classes of R and aR denotes the category in R that contains an element a included in $\Omega$. Let be KB=($\Omega, R$) denotes the knowledge base and B a non empty subset of the set A of all attributes, then the equivalence relation R(B) is called the indiscernibility relation over B representing a binary relation on $\Omega$ defined for x,y $\in \Omega$. Because, information table (relational data) contains attributes and domains, a set $V_a$ is associated to every attribute a ∈ A (its values) and called the domain of a.

Any subset B of A determines a binary relation R(B) on $\Omega$ and is defined as follows:

xR(B)y if and only if a(x)=a(y) for each a ∈ A ,
where a(x) indicates the attribute value a for element x.

Complementary mathematical properties have been explored by the current research in RST. As instance, after studying the ordered set of rough set theory, the author in [10] shows that the relations are not essentially reflexive, symmetric or transitive.

*A. Approximations*

As defined before, as starting point of RST, the indiscernibility relation is intended to express the fact that due to the lack of knowledge, but it is unable to distinguish some objects employing the available information. RST includes another important concept which is Approximations. Approximation is also associated with the meaning of the approximations of topological operations [11].

The types of approximations exploited in Rough Sets Theory are described below:
1. **Lower Approximation ($B_*$):** The description of the domain object known with certainty to belong to the subset of interest defines the lower approximation (LA). Additionally, the LA Set ($B_*$) of a set X regarding to R is the set containing all the objects, which surely can be classified with X regarding R.
2. **Upper Approximation ($B^*$):** The objects that possibly belong to the subset of interest define the upper approximation (UA). Moreover, the UA Set ($B^*$) of a set X with regard to R is the set containing all the objects that, possibly, can be classified with X regarding R.
3. **Boundary Region (BR):** The set of all the objects, contained in a set X with regard to R, which cannot be classified neither as X nor -X regarding R is the definition of BR.

BR is a **crisp set** (exact in relation to R), if the BR is a set $X = \emptyset$ (Empty); otherwise BR is a **rough set** = $B^* - B_*$, if the boundary region is a set $X \neq \emptyset$. More formally, let a set $X \subseteq \Omega$, B be an equivalence relation and a knowledge base $K = (\Omega, B)$. Two subsets can be associated:

1. B-lower: $B_* = \cup \{Y \in \Omega/B : Y \subseteq X\}$
2. B-upper: $B^* = \cup \{Y \in \Omega/B : Y \cap X \neq \emptyset\}$

Similarly, *POS(B)*, *BN(B)* and *NEG(B)* are defined below [8].
3. $POS(B) = B_* \Rightarrow$ certainly member of X
4. $NEG(B) = \Omega - B^* \Rightarrow$ certainly non-member of X
5. $BR(B) = B^* - B_* \Rightarrow$ possibly member of X.

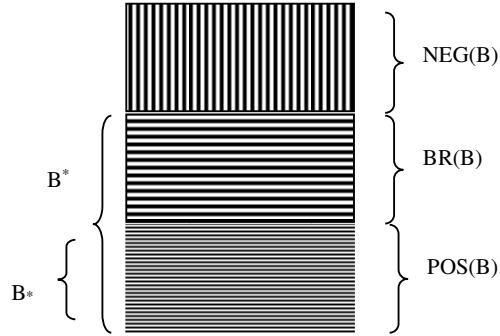

Fig1: B-approximation sets and B-regions Definition

### B. RST Applications

Several properties of RST that make the theory an evident choice for use to deal with real problems: a brief overview of some of the many applications of rough set is presented in the following section:

- **Pattern Recognition:** As an application of pattern recognition, Mrozek and Cyran [12] proposed, in 2001, a hybrid method of automatic diffraction pattern recognition based on RST and Neural Network. This new method uses RST to define the objective function and stochastic evolutionary algorithm for space search of a feature extractor. The neural networks are used for uncertain systems modeling.
- **Acoustical analysis:** An application based on the RST is used to induce generalized rules describing the relationship between acoustical parameters of concert halls and sound processing algorithms is described in the work of Kotek in 1999 [13].
- **Classification of spatial and meteorological pattern:** the current sunspot recognition and classification systems are manual and if successfully learned by a machine, the labor intensive processes begin automated. The approach proposed in [14] by Nguyen et al. in 2005 employs a hierarchical rough set based learning method for sunspot classification. The aim of this system is to learn the modified Zurich classification scheme adopting rough set-based decision tree induction. The evaluation of the proposed system based on sunspots extracted from satellite images, presents promising results. Another work adopting RST approach is developed by Shen&Jensen in 2007 [15] to classify a number of meteorological storm events.
- **Intelligent control systems:** The intelligent control system especially when incorporated with fuzzy theory is an important application field of rough set theory [16].

## IV. RST APPLIED TO CAR

### A. Data representation with RST

The format, often, used to present data is table format, where each column indicates an *attribute* and each row indicates an *object* of interest and each entry of the table contains an *attribute value*. Such tables are composed of *information systems, attribute-value tables* and *information tables*. In this paper, we will adopt the information table format, where the columns represent variables and rows represents cases (objects). All variables in information tables are called attributes.

The main problems that can be undertaken by the use of RST are the following:
- A set of object can be characterized in terms of attribute values.
- It is possible to find association rules between items in *Y* and *I*.
- Generation of association rules

An example of information table is presented in Table 5 with two classes *Y*={Sport and Education} and seven text documents. Each document is a transaction and consists of a set of keywords. Additionally, each transaction is labelled with a topic class in *Y*. The set of keywords is denoted by the items in *I*={*Student, Teach, School, City, Game, **Baseball, Basketball, Team, Coach, Player, Spectator***}.

TABLE VI
Example of illustrative data set containing documents and their classes.

| Doc id | Transaction | Class |
|---|---|---|
| 1 | Student, Teach, School | Education |
| 2 | Student, School | Education |
| 3 | Teach, School, City, Game | Education |
| 4 | Baseball, Basketball | Sport |
| 5 | Basketball, Player, Spectator | Sport |
| 6 | Baseball, Coach, Game, Team | Sport |
| 7 | Basketball, Team, City, Game | Sport |

The set $\Omega$ represents all the possible cases, the set of all attributes denoted by *A*, and the set of all attribute values denoted by V. An information table defines an information function I: $\Omega \times A \rightarrow V$.

Pawlak have presented a formal definition of a decision table, in 1982. A decision table is a system S= ($\Omega$, A, V, f) where:
- A constitutes the union of the conditions attributes set (C) and the decision attributes set (D) (C$\cup$D)
- V: denotes the union of the set of values of an attribute a included in A (domain of a) represented as follows:

$$\bigcup_{a \in A} V_a$$

- fa: is an association rule function between attributes fa: $C_a \rightarrow D_a$, Where $C_a \subseteq$ C an attribute or a set of attributes that belongs to C and $D_a \subseteq$ D an attribute or a set of attributes that belongs to D. The association rule is denoted by a function fv:Cv$\rightarrow$Dv, Where Cv$\subseteq$ C a value or a set of values that belongs to Cv and Dv$\subseteq$ D an attribute or a set of attributes that belongs to Dv.
- Table 6 contains attributes, where condition attributes are in the set ={Student, Teach, School, City, Game, Baseball, Basketball, Team, Coach, Player, Spectator} and decision attribute in the set {class}.

An attribute-value is denoted by the pair $\tau = (a, v)$ where $a \in A$, $v \in V$. $[\tau]$ denotes a block, including the set of all cases in $\Omega$ where each attribute *a* has a value *v*. In ARM approach, the support measure of an attribute, compute the existence of an attribute in a specified row, then the support of an attribute-value pair is obtained by the cardinality of $[\tau]$ and denoted by $|[\tau]|$. Based on the example in the Table 6, blocks and their related support are defined as follows:

$[\tau]_1$: [{Student}] = {1, 2}, and support($[\tau]_1$)=2
$[\tau]_2$:[ {School}] = {1,2,3}, and support($[\tau]_2$)=3
$[\tau]_3$:[ {Spectator}] = {5}, and support($[\tau]_3$)=1
$[\tau]_4$:[ {Basketball}] = {4, 5, 7}, and support ($[\tau]_4$)=3
$[\tau]_5$=[{Game}] = {3,6, 7}, and support ($[\tau]_5$)=3
$[\tau]_6$=[ {Baseball}] = {4,6}, and support ($[\tau]_6$)=2
$[\tau]_7$=[{Student, School}] = {1, 2}, and support($[\tau]_7$)=2
$[\tau]_8$=[{Team}] = {6, 7}, and support($[\tau]_8$)=2

Let be x∈ $\Omega$ and B ⊆ A. We denote the elementary set of B containing *x* by $[x]_B$, representing by the following set: $\cap\{[(a, v)] \mid a \in B, I(x, a) = v\}$

Let be the subset of $\Omega$ containing all cases from $\Omega$ that are indistinguishable from *x* while using all attributes from *B* the elementary sets. Elementary sets are called *information granules* in the terminology of *soft computing*. Element sets are blocks of attribute-value pairs represented by that specific attribute, while subset *B* is limited to a single attribute, Consequently,

[{Game}]={3,6,7}
[{Player}]={ 5}

To combine two attribute-values, for example, the elementary set of B with two attributes is defined as follows:

- $\{[\tau]_1,[\tau]_2\}$=[{Student, School}]={1,2}, and support $([\tau]_1,[\tau]_2)$=2
- $\{[\tau]_5, [\tau]_8\}$=[{Game,Team}]={6,7}, and support $([\tau]_5, [\tau]_8)$=2

## B. Class association rules Algorithm

### B.1. Class association rules between items

CARs can be mined directly in a single step, unlike the normal association rules. The aim is to find all rules having a support greater than ***minsupp***, and for that reason a rule is of the form: (*i, y*) where i⊆ $I$ (set of items) and y⊆ $Y$ (a class label).

The support and the confidence of a class association rules are denoted, respectively, by S and C as follows:

$$S = \frac{|B_*(i) \cup B_*(y)|}{|\Omega|}$$

Where B* is the upper approximation in term of rough set theory representing the items in the condition of the rule and $|B_*(i) \cup B_*(y)|$ the number of the items i occurring in conjunction with a label y across the transactions in the table and $|\Omega|$ indicates the number of all the transactions in the table.

$$C = \frac{|B_*(i) \cup B_*(y)|}{|B_*(i)|}$$

Where $B_*(i)$ denotes the number of the items i in the condition of the association occurring across the transactions in the table.

Let be a class association rule defined as follows: **CR**={Student, School➔Education}.
The elementary set of **B** in the condition of the rule contains two attributes and is defined as follows:

- {condSet}=$\{[\tau]_c\}$=[{Student, School}]={1,2}, and support $([\tau]_c)$=2
- {decSet}={ $[\tau]_d$}=[{education}]={1,2,3} and support { $[\tau]_d$}=3
- Support of CR= support {condSet∪decSet} =support$\{[\tau]_c,[\tau]_d\}$=support{1,2}=2
- $|\Omega| = 7$

Then the support of (CR) is 2/7=28%.
The confidence of CR is the S(CR)/support ($[\tau]_c$)=2/2=1.
However, as these explained by the previous examples, the rough set approach to discover CAR is much simpler than the normal association method presented in the beginning of this paper.

### B.2. CAR mining Algorithm

The algorithm generating class association rules is denoted by C_Apripori which is based on Apriori algorithm. C_Apriori generates all the frequent rules making multiple passes over data resembling the Apriori algorithm. In the first pass, it counts the support of each 1-ruleitem (containing one item in its condition set). The set of all ruleitems (1-candidate) is denoted by the following expression:
$C_0$={(({i},y)|i⊆ $I$ and y⊆ $Y$}

**Algorithm C_Apriori**

```
1    Discretization of data, k=0;
2    C_k ← init ( ) ;              //first pass over database
3    F_k ← {f|f∈C_0, f.support≥minsupp};
4    CR_k ← {f|f⊆F_k , f.confidence≥minconf} ; k++
5    for (i=k ; F_{k-1} ≠ ∅ ; i++) do
6      C_i ← CAcandidate-gen(F_{i-1});
7       for each transaction t⊆T do
8        for each c⊆C_i do
9         if (c.Condset is included in t) then
10            c.condsupport++
11        if(t.class=c.class) then
12            CR_i.support++
13       endfor
14      endfor
15      F_i ← {c∈C_i|c.support≥minsupp} ;
16      CA_i ← {f|f⊆F, f.support≥minconf} ;
17   endfor
18   return CA ← ∪_i CA_i
```

The instruction in line 3 indicates whether the candidate 1-ruleitems are frequent or no and we generate 1-condition CR (rule with unique condition) from the identified 1-ruleitem. In the next pass i, the algorithm C_Apriori starts with the beginning set of (i-1)-ruleitems established as frequent in the (i-1)-pass, and uses this beginning set to generate other new frequent k-ruleitems ($C_i$ in line 6). The support counted for both the condition rule and the rule are updated continuously during the scan of the data for each i-ruleitem. The objective behind the overall data scan is to find which of the actually frequent candidate k-ruleitem in $C_i$ (line 15). And finally, in line 16, the C_Apriori algorithm generates i-condition CA (class association rules with i conditions). The CAcandidate-gen is very similar function to the candidtae-gen function in the Apriori algorithm. The unique difference is that in CAcandidate-gen ruleitems joins the condition sets aiming to join the ruleitems with same class.

## V.  CONCLUSION

This paper proposes an approach based RST for class association rule mining. Mining class association rules with the proposed C_Apriori algorithm is easy and efficient. It computes the support and the confidence in a similar manner to the elementary set of lower approximation included in the RST approach. C_Apriori is more easily compared to the classic Apriori algorithm, where the process of frequent itemsets searching based on the concept of equivalence class is very simple. In future we will investigate the Bitmap structure to convert dataset to structured data where items are denoted by binary representation, and each line (transaction) is converted to a binary number.

## AUTHOR PROFILE

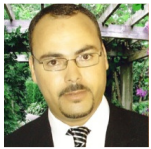

**Dr. Thabet Slimani** got a PhD in Computer Science (2011) from the University of Tunisia. He is currently an Assistant Professor of Information Technology at the Department of Computer Science of Taif University at Saudia Arabia, where he is involved both in research and teaching activities. His research interests are mainly related to Semantic Web, Data Mining, Business Intelligence, Knowledge Management and recently Web services. Dr.Thabet is the author of some programming books and has published his research through international conferences, chapter in books and peer reviewed journals. He also serves as a reviewer for some conferences and journals.